\documentclass[preprint,12pt]{elsarticle}
\journal{Chaos Solitons and Fractals}
\usepackage{graphics}      
\usepackage{graphicx}      
\usepackage{longtable}     
\usepackage{url}           
\usepackage{bm}            
\usepackage[usenames]{color}
\usepackage[dvipsnames]{xcolor}
\usepackage{cancel}
\usepackage{soul}
\usepackage{amsmath,amssymb}
\usepackage{epstopdf}
\usepackage{mathtools}
\usepackage{natbib}
\usepackage{amsfonts}
\usepackage{amsthm,mathrsfs,amsopn}
\usepackage[utf8]{inputenc}
\usepackage{tikz}
\usepackage{siunitx}        
\usepackage[french]{babel}  
\usepackage{icomma}
\theoremstyle{plain}

\begin{document}

\title{Chimera state in neural network with the Proportional-Integral-Derivative coupling}
\author[a,b]{Adamdine M. Abdoulaye }
\affiliation[a]{Research Unit Condensed Matter, Electronics and Signal Processing,
University of Dschang, P.O. Box 67 Dschang, Cameroon.}
\affiliation[b]{MoCLiS Research Group, Dschang, Cameroon}

\author[a,b]{Venceslas Nguefoue Meli}%

\author[a,b]{Steve J. Kongni}%

\author[c,d,b]{Thierry Njougouo\corref{cor}}
\affiliation[c]{IMT School for Advanced Studies, Lucca, Italy}
\affiliation[d]{Faculty of Computer Science and naXys Institute, University of Namur, Namur, Belgium}

\cortext[cor]{ Corresponding author: IMT School for Advanced Studies, Lucca, Italy}
\ead{thierry.njougouo@imtlucca.it}

\author[a,b,e]{Patrick Louodop}
\affiliation[e]{Potsdam Institute for Climate Impact Research (PIK) Member of the Leibniz Association P.O. Box 60 12 03 D-14412 Potsdam Germany}

\date{\today}

\begin{abstract}
This study delves into the emergence of collective behaviors within a network comprising interacting cells. Each cell integrates a fixed number of neurons governed by an activation gradient based on Hopfield's model. The intra-cell interactions among neurons are local and directed, while inter-cell connections are facilitated through a PID (Proportional-Integral-Derivative) coupling mechanism. This coupling introduces an adaptable environmental variable, influencing the network dynamics significantly.
Numerical simulations employing three neurons per cell across a network of fifty cells reveal diverse dynamics, including incoherence, coherence, synchronization, chimera states, and traveling wave. These phenomena are quantitatively assessed using statistical measures such as the order parameter, strength of incoherence, and discontinuity measure. Variations of the resistive, inductive, or capacitive couplings of the inter-cell environment are explored and their effects are analysed.
Furthermore, the study identifies multistability in network dynamics, characterized by the coexistence of multiple stable states for the same set of parameters but with different initial conditions. A linear augmentation strategy is employed for its control. 
\end{abstract}


\maketitle
\section{Introduction}
The study of complex systems and nonlinear dynamics has significantly enriched our understanding of living systems and their applications in various fields, including human life and technology. For instance, in biology and medicine, nonlinear dynamics are essential for understanding circadian rhythms and sleep disorders\cite{golombek2010physiology}, as well as for modeling the spread of infectious diseases to develop effective containment strategies and controls\cite{anderson1991infectious,struelens2024real}. In neuroscience, these models aid in understanding neural networks and brain dynamics associated with diseases like epilepsy, Alzheimer's disease, and Parkinson's disease, as well as exploring the mechanisms underlying various states of consciousness\cite{jiruska2013synchronization,tononi1998consciousness}.
In the field of technological, the principles of complex systems optimize communication networks and inspire collective robotics, which are useful in space exploration and disaster management\cite{solgi2021bee,adams2023self,antonic2024collective}.

The application of complex networks in neural systems has revolutionized our understanding of brain function and neurological disorders. By modeling the interactions between neurons as a network, researchers have developed mathematical frameworks that capture the intricate dynamics of neural activity. Investigations on networks have shown that collective behaviors arise from several factors such as topology \cite{gomez2007synchronizability}, coupling strengths \cite{rogers1994ordering}, or the nature of the coupling functions \cite{stankovski2017coupling}. These models, such as the Hodgkin-Huxley \cite{hodgkin1952quantitative}, FitzHugh-Nagumo \cite{rocsoreanu2012fitzhugh}, Hindmarsh-Rose \cite{storace2008hindmarsh}, and Hopfield \cite{farhat1985optical} models, simulate the electrical and chemical processes within and between neurons, providing insights into synchronization phenomena and the emergence of complex states like chimera, multichimera, travelling chimera states \cite{mitchell1997artificial,yegnanarayana2009artificial,abraham2005artificial,simo2021chimera}. These states, characterized by the coexistence of synchronized and desynchronized regions, are not only theoretical constructs but have also been observed in various physical and biological systems, highlighting their relevance to understanding brain disorders \cite{bansal2019cognitive,calim2020chimera,masoliver2022embedded,simo2021chimera}.

Furthermore, advancements in synaptic modeling, incorporating resistive, capacitive, and inductive properties of the neuronal environment, have enhanced our ability to simulate and control neural dynamics \cite{bahramian2021collective,zhou2021synaptic,li2021simulation}. This has significant implications for developing treatments for neurological diseases, brain-computer interfaces, and artificial intelligence systems that mimic brain-like processing \cite{iqbal2017modeling}. PID (Proportional-Integral-Derivative) coupling is a widely utilized control methodology in various dynamic systems, including neuronal networks, to enhance stability and performance. This control strategy is instrumental in managing the complex interactions between neurons by incorporating the electrical, chemical, and magnetic properties of synapses. The PID controller continuously adjusts the system's input based on the current error, the integral of past errors, and the derivative of future error predictions. This dynamic adjustment facilitates the synchronization of neuronal activity and the emergence of complex behaviors such as chimera states and multichimera states. Studies indicate that the application of PID coupling in neuronal networks can lead to a deeper understanding of brain disorders and significant advancements in brain-machine interfaces and artificial intelligence systems\cite{ang2005pid,liu2001optimal,borase2021review,mcmillan2012industrial,taguchi2002tuning,wang2021simulation,rehman2017control,jia2013modeling,kang2014adaptive,wang2015fuzzy}. By integrating PID control, neuronal networks can become more resilient to disturbances and better mimic observed biological behaviors.

In this paper, we study a neural network composed of N cells non-locally coupled and each cell is formed by three Hopfield neurons. The cells are connected through a single intra-cell neuron via a PID controller, forming a kind of network of network. Each small network corresponds to a cell with a fixed number of neurons ($n=3$). The objective is to investigate the dynamics of the entire network as a function of the parameters of the PID controller. Given that the dynamics of coupled systems are highly dependent on their initial conditions, we aim to control the multistability using a linear augmentation strategy \citep{njitacke2021window}. Multistability is a remarkable property of nonlinear interactive systems, where a system can reside in multiple stable states depending on initial conditions \cite{feudel1997multistability,sharma2014controlling,sharma2015control}. This characteristic is crucial for understanding various natural and pathological phenomena, such as cell cycles and immune memory. Controlling multistability can lead to significant advancements in managing complex system behaviors, potentially providing new insights into neuronal diseases and improving technological applications involving neural networks. Through our investigation, we aim to elucidate the intricate dynamics and control mechanisms of such systems, highlighting their potential applications in neuroscience and beyond.

The remainder of this paper is organized as follows: Section \ref{sec2} introduces the mathematical model that underpins our study, detailing the formulation of the neural network consisting of locally coupled Hopfield neurons and the incorporation of PID coupling mechanisms. Section \ref{sec3} is dedicated to presenting the numerical results obtained from our simulations. Here, we delve into the effects of varying PID coupling parameters on the collective behaviors observed within the network. By systematically varying these parameters and analyzing their impact on synchronization, chimera states, and potentially multistable behaviors, we aim to provide insights into how specific configurations influence network dynamics. Finally, in Section \ref{sec5}, we synthesize our findings and draw conclusions from the study.

\section{Model} \label{sec2}
\begin{figure}[htp!]
\centering
  \begin{tabular}{cc}
    \includegraphics[width=0.4\textwidth]{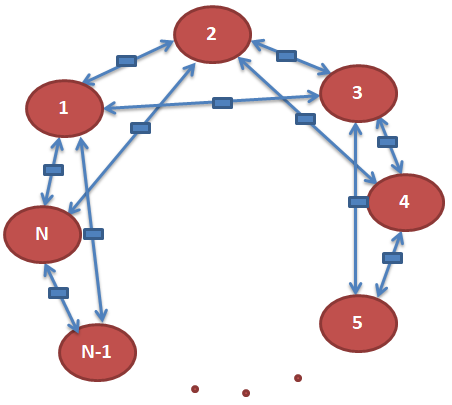} & 
    \includegraphics[width=0.6\textwidth]{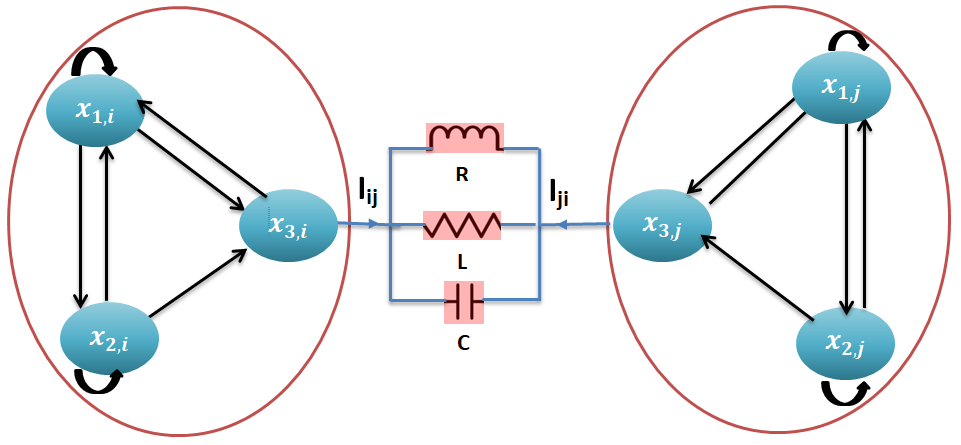} \\ (a) & (b)
\end{tabular}
\caption{Scheme of non-local coupling of a network of $N$ cells: (a) topology and (b) expanded model of a cell with PID coupling between two cells $i$ and $j$.}\label{topol}
\end{figure}

The model considered for this study is a network of $N$ cells interacting with non-local and undirected coupling, as illustrated in Fig.\ref{topol}(a). Each cell in this network is a model of a small neural network composed of $n=3$ (three) Hopfield neurons connected by gradient activation, as proposed in Refs.\cite{doubla2020effects, njitacke2020extremely} (see Fig.\ref{topol}(b) for the detailed representation of two cells and their connections). 
To analyze the interactions between neurons within each cell of the network, we use the $n \times n$ weighted and directed connectivity matrix in Eq.\ref{eq2}. 

\begin{equation}\label{eq2}
w = \left( {\begin{array}{*{20}{c}}
2&{ - 1.2}&{0.48}\\
{3.6}&{1.7}&{1.076}\\
{ - 9}&0&0
\end{array}} \right)
\end{equation}

Let us describe the dynamics of the neurons in each cell of the network using Eq.\ref{eq1}:
\begin{equation}\label{eq1}
  {\dot x_k} =  - {x_k} + \epsilon \sum\limits_{\ell = 1}^n {{w_{k\ell}}\tanh \left( {{\beta _\ell}{x_\ell}} \right)}
\end{equation}
where, $x$ represents the state vector of the neurons, $k = 1, 2, 3$ is the index of the neuron within each cell and $\epsilon$ the intra-cells coupling strength. Neuron activation is modeled using the hyperbolic tangent function $(\tanh)$, which features an adjustable gradient. The parameter $\beta$ serves as a control variable, determining the slope of the function. 

The coupling between the two cells, depicted as rectangles in Fig.\ref{topol}(a), is managed through a PID (Proportional, Integral, and Derivative) controller, despite the nonlocal nature of their interaction. This control mechanism ensures precise regulation by combining proportional response, cumulative history, and rate of change to achieve effective coordination between the cells. This circuit, consisting of a resistor, capacitor, and inductor all connected in parallel, is illustrated in Fig.\ref{topol}(b).

From Eq.\ref{eq1}, we propose the mathematical model applicable to the $N$ cells of the network depicted in Fig.\ref{topol}(a). Considering the circuit and network topology, the coupling between two cells is represented by the electrical current, denoted $I_{ij}$, which signifies the flow of signal from node $i$ to node $j$. Hence, the mathematical model describing the entire network is given by Eq.\ref{eq3}.

\begin{equation}\label{eq3}
  \dot x_{k,i} =  - x_{k,i} + \epsilon \sum\limits_{\ell = 1}^n {{w_{k\ell}}\tanh \left( {{\beta _k}x_{\ell,i}} \right)}  + {I_{ij}}
\end{equation}

Let $x_{k,i}$ represent the \(k\)-th neuron within the cell indexed by \(i\). In this investigation, we focus on the coupling between the cells specifically at the third neuron of each cell. However, this approach can also be applied to any other neuron within the cells, such as neuron 1 or 2 (see Fig.\ref{topol}(b)). We model the coupling between two cells \(i\) and \(j\) (intercellular coupling) using the additional term \(I_{ij}\) in Eq.\ref{eq3}. For simplicity, we omit the neuron index within the cell, as we assume that the intercellular coupling occurs through the third neuron of each cell. This additional term represents the current flowing through a parallel RLC circuit, which acts as a PID coupling mechanism. The properties of this circuit integrate the electrochemical processes characterizing the inter-neuronal environment \cite{rehman2017control, wang2015fuzzy}.
Hence, the mathematical expression of $I_{ij}$ for this scenario of nonlocal coupling with $2q$ nearest neighbors is provided by Eq.\ref{eq4}.
\begin{equation}\label{eq4}
  \begin{array}{l}
{I_{ij}} = \frac{{{\sigma_P}}}{{2q}}\sum\limits_{j = i - q}^{ i + q} {\left( {x_{3,j} - x_{3,i}} \right)}  + \frac{{{\sigma_D}}}{{2q}}\sum\limits_{j = i - q}^{i + q} {\left( {\dot x_{3,j} - \dot x_{3,i}} \right)} + \frac{{{\sigma_I}}}{{2q}}\sum\limits_{j = i - q}^{i + q} {\int {\left( {x_{3,j} - x_{3,i}} \right)dt} }
\end{array}
\end{equation}
In Eq.\ref{eq4}, the terms ${\sigma_P} = \frac{1}{R}$, ${\sigma_D} = C$, and ${\sigma_I} = \frac{1}{L}$ represent the resistive, capacitive, and inductive coupling coefficients, respectively. The term $\left({x_{3,j} - x_{3,i}}\right)$ in Eq.\ref{eq4} represents the potential difference between the third neuron of the $i^{th}$ cell and the $j^{th}$ cell at time $t$. Just as a reminder, the variables $x_{k,i}$ and $I_{ij}$ are implicitly functions of time, and for the sake of simplicity in equations, we chose to omit the explicit time dependence.

To simplify the numerical solution process, we propose converting this system into a set of ordinary differential equations (ODEs). By differentiating the  variable \({\dot{x}}_{3,i}\) in Eq.\ref{eq3} and subsequently simplifying it, we derive 
Eq.\ref{eq4aa2} as shown below:

\begin{equation}\label{eq4aa2}
\begin{array}{l}
{{\ddot x}_{3,i}}\left[ {1 + {\sigma _D}} \right] - \frac{{{\sigma _D}}}{{2q}}\sum\limits_{j = i - q}^{j = i + q} {{{\ddot x}_{3,j}}}  =  - {x_{3,i}} - \frac{{9{\beta _1}}}{{{{\left( {\cosh \left( {{\beta _1},{x_{1,i}}} \right)} \right)}^2}}}\\
\,\,\,\,\,\,\,\,\,\,\,\,\,\,\,\,\,\,\,\,\,\,\,\,\,\,\,\,\,\,\,\,\,\,\,\,\,\,\,\,\,\,\,\,\,\,\,\,\,\,\,\,\,\,\,\,\,\,\, + \frac{1}{{2q}}\sum\limits_{j = i - q}^{j = i + q} {\left[ {{\sigma _P}\left( {{{\dot x}_{3,j}} - {{\dot x}_{3,i}}} \right) + {\sigma _I}\left( {{x_{3,j}} - {x_{3,i}}} \right)} \right]} 
\end{array}
\end{equation}

By introducing a fourth variable \(x_{4,i}\), defined such that \(\dot{x}_{3,i} = x_{4,i}\), into the system of ordinary differential equations, the degrees of freedom for each cell in the system are increased. Consequently, Eq.\ref{eq4aa2} is transformed as follows.

\begin{equation}\label{eq4aa}
{\dot x_{4,i}}\left[ {1 + {\sigma _D}} \right] - \frac{{{\sigma _D}}}{{2q}}\sum\limits_{j = i - q}^{j = i + q} {{{\dot x}_{4,j}}}  = {F_i}\left( X \right)
\end{equation}

The analytical expression of the function ${F_i}(.)$ is given by Eq.\ref{eq6}.
\begin{equation}\label{eq6}
  \begin{array}{l}
{F_i} =  - x_{4,i} - \frac{{9\epsilon {\beta _1}\dot x_{1,i}}}{{{{\left( {\cosh \left( {{\beta _1}x_{1,i}} \right)} \right)}^2}}} + \frac{1}{{2q}}\sum\limits_{l = i - q,\,i \ne l}^{i + q} {\left( {{\sigma_P}\left( {x_{4,\ell} - x_{4,i}} \right) + {\sigma_I}\left( {x_{3,\ell} - x_{3,i}} \right)} \right)}
\end{array}
\end{equation}
By combining these equations, the network is defined by the set of the following ODE.

\begin{equation}\label{eq5}
  \left\{ \begin{array}{l}
\dot x_{1,i} =  -  x_{1,i} + \epsilon (2\tanh \left( {{\beta _1}x_{1,i}} \right) - 1.2\tanh \left( {{\beta _2}x_{2,i}} \right) + 0.48\tanh \left( {{\beta _3}x_{3,i}}\right))\\
\dot x_{2,i} =  - x_{2,i} + \epsilon (3.6\tanh \left( {{\beta _1}x_{1,i}} \right) + 1.7\tanh \left( {{\beta _2}x_{2,i}} \right) + 1.076\tanh \left( {{\beta _3}x_{3,i}} \right))\\
\dot x_{3,i} = x_{4,i}\\
\dot x_{4,i} = \sum\limits_j {M_{ij}^{ - 1}{F_j}}
\end{array} \right.\,
\end{equation}
where M is a matrix as defined by Eq.\ref{eq7}.

\begin{equation}\label{eq7}
  \begin{array}{l}
M = \left[ {\begin{array}{*{20}{c}}
\lambda &{  \gamma }&{  \gamma }&0&.&.&.&0&0&{  \gamma }&{  \gamma }\\
{  \gamma }&\lambda &{  \gamma }&{  \gamma }&{}&{}&{}&{}&0&0&{  \gamma }\\
{  \gamma }&{  \gamma }&\lambda &{  \gamma }&{}&.&{}&{}&{}&0&0\\
0&{  \gamma }&{  \gamma }&\lambda &{}&.&.&{}&{}&{}&0\\
.&{}&{}&{}&{}&.&.&.&{}&{}&.\\
.&{}&.&.&.&.&.&.&{}&{}&.\\
.&{}&{}&.&.&.&.&.&{}&{}&.\\
0&{}&{}&{}&.&.&.&.&{}&{}&0\\
0&0&{}&{}&{}&{}&{}&{}&{}&{}&{  \gamma }\\
{  \gamma }&0&0&{}&{}&{}&{}&{}&{}&{}&{  \gamma }\\
{  \gamma }&{  \gamma }&0&0&.&.&.&0&{  \gamma }&{  \gamma }&\lambda
\end{array}} \right]\\
\end{array}
\end{equation}
with $\lambda  = 1 + {\sigma_D}$ and $\gamma  = -\frac{{{\sigma_D}}}{{2q}}$\\
In the next section, we will explore the different patterns that can be obtained from this model.

\section{Numerical results}\label{sec3}
\subsection{Chimera states}
In this section, we explore the collective behaviors emerging in this network, described mathematically by the set of Eq.\ref{eq5}. For the numerical simulation, we use the 4th-order Runge-Kutta method over $50,000$ iterations with a time step of $dt = 0.02$, and the solution is saved after discarding $70\%$ of the simulation time as the transient phase. Various collective behaviors can be observed, such as synchronization, chimera states, coherent states, etc. To analyze these behaviors, we employ a range of statistical techniques. First, we employ the order parameter introduced by Kuramoto and Battogtokh \cite{kuramoto2002coexistence}, which is effective in discerning phase synchronization and non phase synchronization (see \ref{appA} for description). Additionally, we apply the Strength of Incoherence (SI) and the Discontinuity Measure (DM) developed by Gopal et al.\cite{gopal2014observation}. These metrics help differentiate between incoherent, chimera, and coherent states within networks (see \ref{appB} for details).

\begin{figure}[htp!]
\centering
  \begin{tabular}{cc}
    \includegraphics[width=0.5\textwidth]{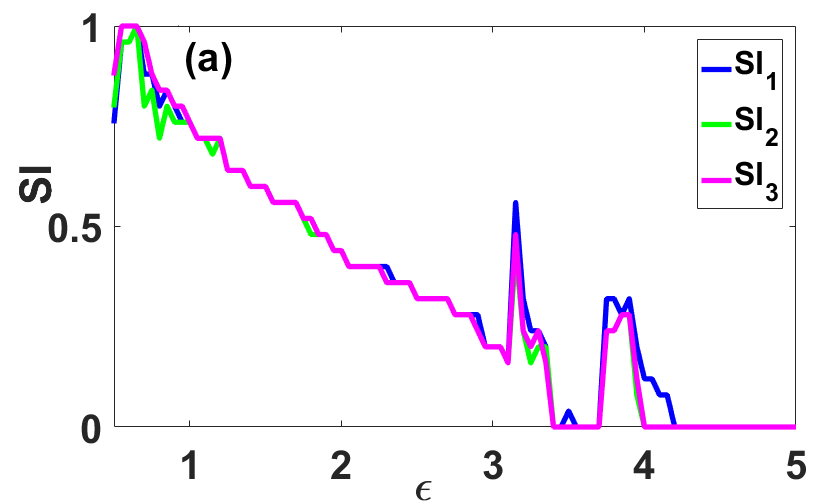}&
    \includegraphics[width=0.5\textwidth]{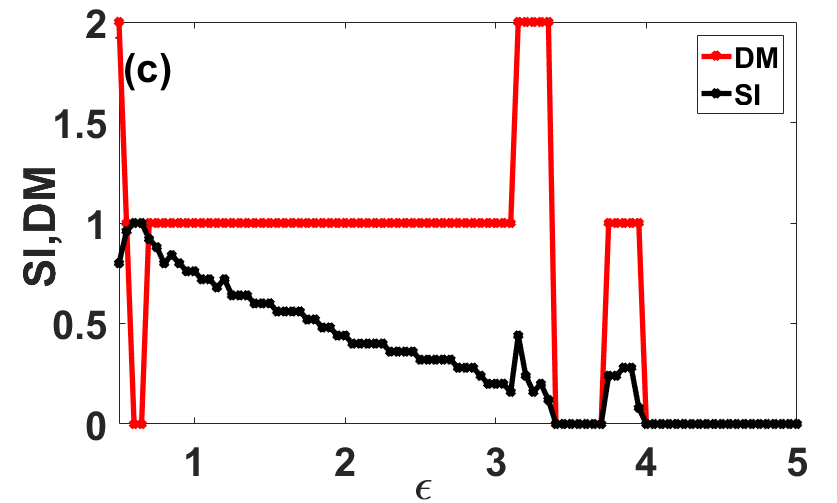}\\
    \includegraphics[width=0.5\textwidth]{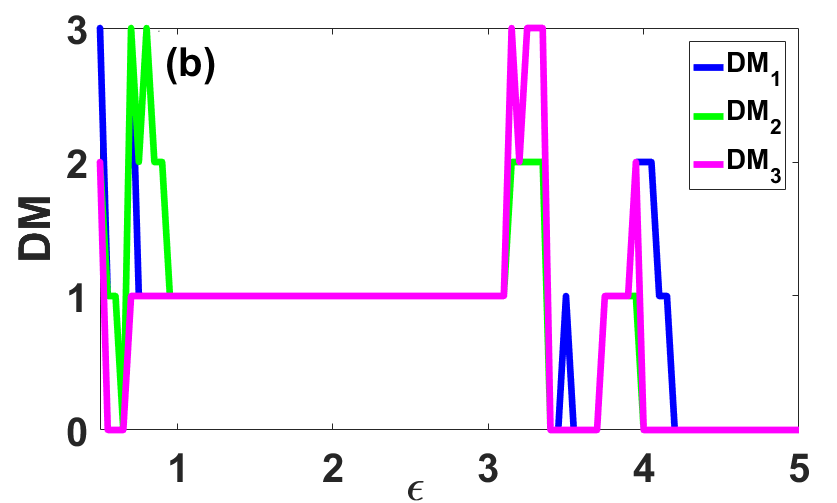}&
    \includegraphics[width=0.5\textwidth]{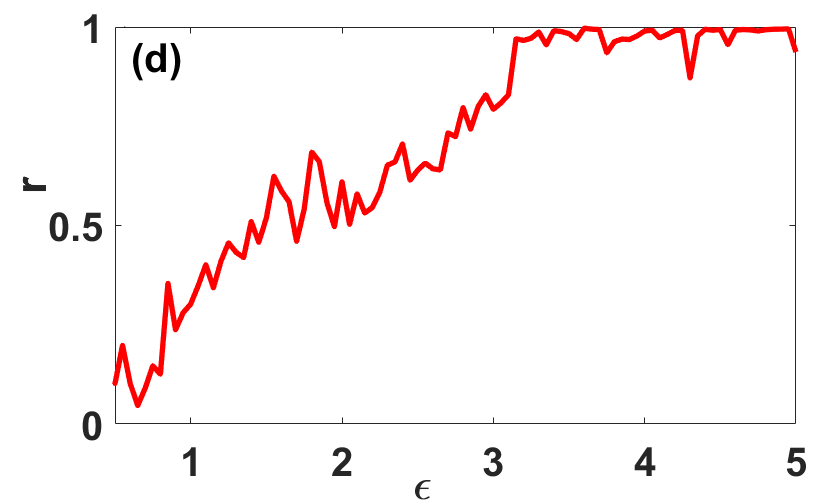}
\end{tabular}
    \caption{Characterization of the chimera states using the $SI$ and $DM$ and phase synchronization according to the intra-cell coupling strength $\epsilon$. (a) $SI_k$ for each of the $k$ neurons in each cell of the network; (b) $DM_k$ for each of the $k$ neurons in each cell of the network. The color blue corresponds to the first neurons $(k=1)$, green to the second neurons $(k=2)$, and magenta to the third neurons $(k=3)$. (c) Average $SI$ and $DM$ for the $k$ neurons in each cell of the network and (d) average order parameter $r$ obtained using the mean of the phase in the cells. The other parameter values are: $\sigma_D=0.03$, $\sigma_P=1.54$, and $\sigma_I=0.01$.
     } \label{oder1}
\end{figure}

For the rest of this work, we assume the parameters $\beta_1, \beta_2$, and $\beta_3$ for the three neurons forming each Hopfield neuron in cell are all identical and equal to one. We also fix the number of cells to $N = 50$ and the number of nearest neighbors cell to $q = 2$. With the inter-cell coupling parameters set to \(\sigma_D = 0.03\), \(\sigma_P = 1.54\), and \(\sigma_I = 0.01\), Fig.\ref{oder1} illustrates the dynamics of the Strength of Incoherence (SI), the Discontinuity Measure (DM), and the order parameter \( r \) as functions of the intra-cell coupling strength \(\epsilon\). The values of SI and DM are computed separately for each neuron \( k \) within a cell, labeled as \( SI_k \) and \( DM_k \) for \( k = 1, 2, 3 \). Fig.\ref{oder1}(a) and \ref{oder1}(b) display the results for \( SI_k \) and \( DM_k \), with blue, green, and magenta colors representing the first (\(k=1\)), second (\(k=2\)), and third neurons (\(k=3\)), respectively. The evolution patterns of \( SI_k \) and \( DM_k \) reveal close alignment dynamics across all neurons in each cell. To provide a more streamlined view, Fig.\ref{oder1}(c) presents the average SI and DM values across the three neurons per cell.
These results reveal that, with the PID inter-cell coupling coefficients considered, the intra-cell coupling strength \(\epsilon\) drives a gradual transition of the system from full decoherence to total coherence. Specifically, as \(\epsilon\) varies, the system progresses from a state of decoherence (\(SI = 1\) and \(DM = 0\)) to complete coherence (\(SI = 0\) and \(DM = 0\)), passing through intermediate states. In these intermediate regimes, we observe chimera states, characterized by partial coherence (\(0 < SI < 1\) and \(DM = 1\)), as well as multichimera states, where coherence is more complexly structured (\(0 < SI < 1\) and \(DM \geq 2\)). Given the limitations of $SI$ and $DM$ in fully characterizing synchronization, the use of the order parameter $r$ (see Fig.\ref{oder1}(d)) allows us to demonstrate that the coherence predicted by these tools corresponds to a state of phase synchronization. This is evident as $r \to 1$ for $\epsilon > 4$. Therefore, this transition towards synchronization is progressive, passing through various collective behaviors.

\begin{figure}[htp!]
\centering
  \begin{tabular}{cc}
    \includegraphics[width=0.5\textwidth]{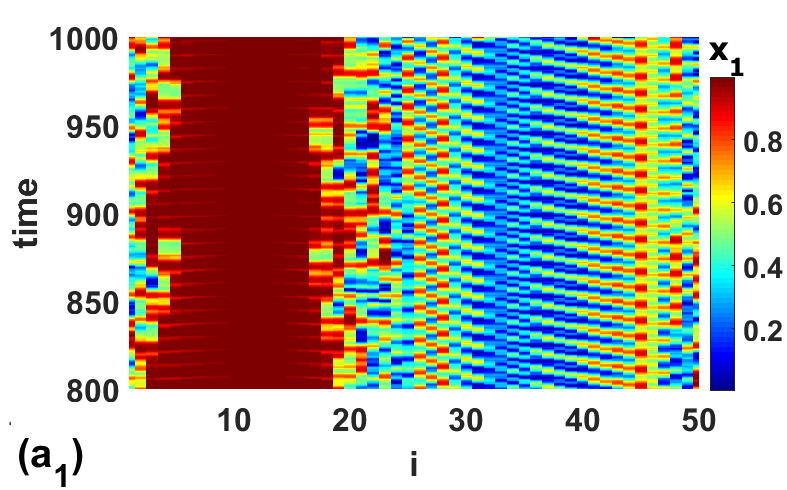}&
    \includegraphics[width=0.5\textwidth]{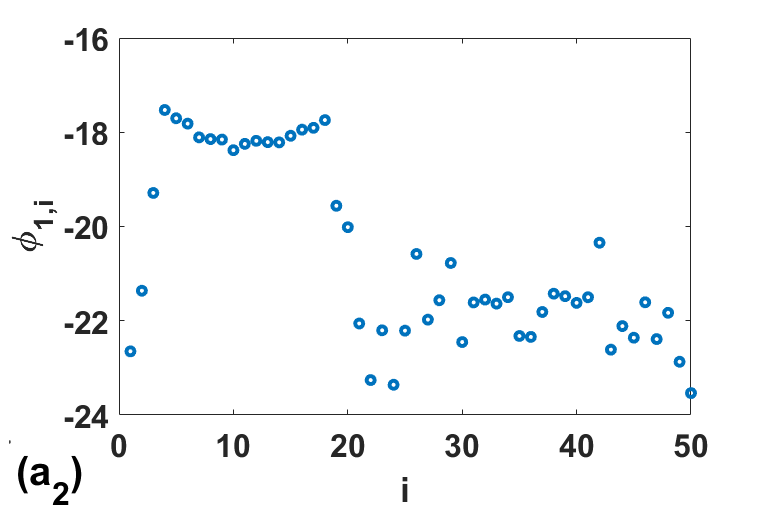}\\
    \includegraphics[width=0.5\textwidth]{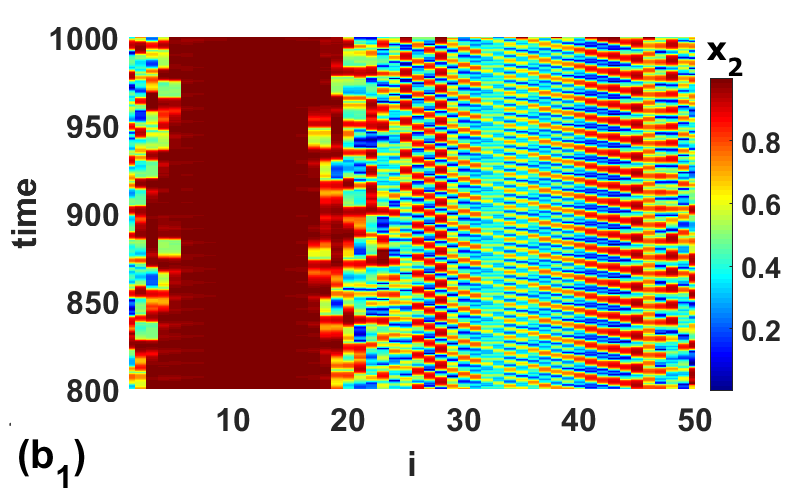}&
    \includegraphics[width=0.5\textwidth]{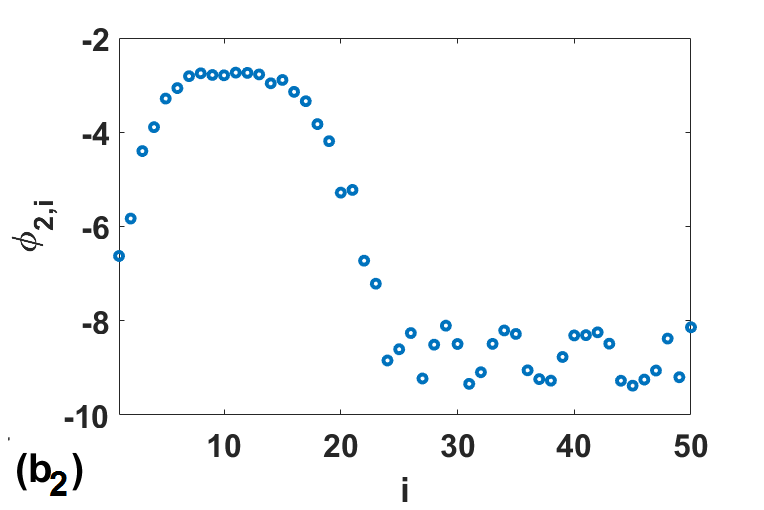}\\
    \includegraphics[width=0.5\textwidth]{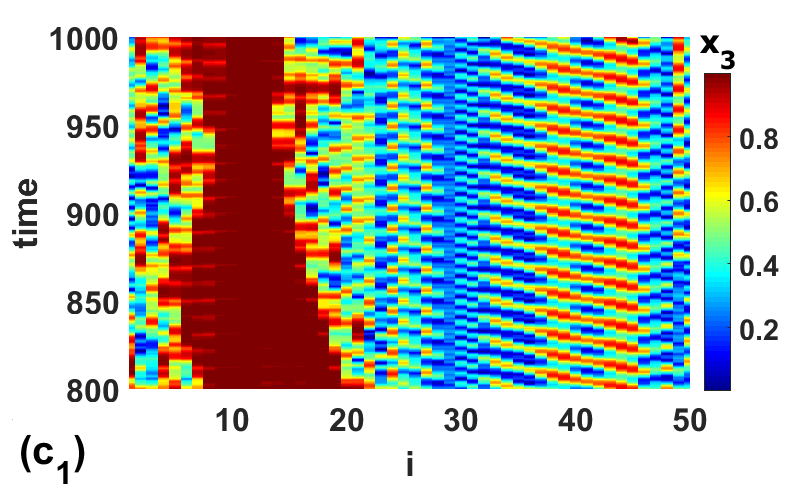}&
    \includegraphics[width=0.5\textwidth]{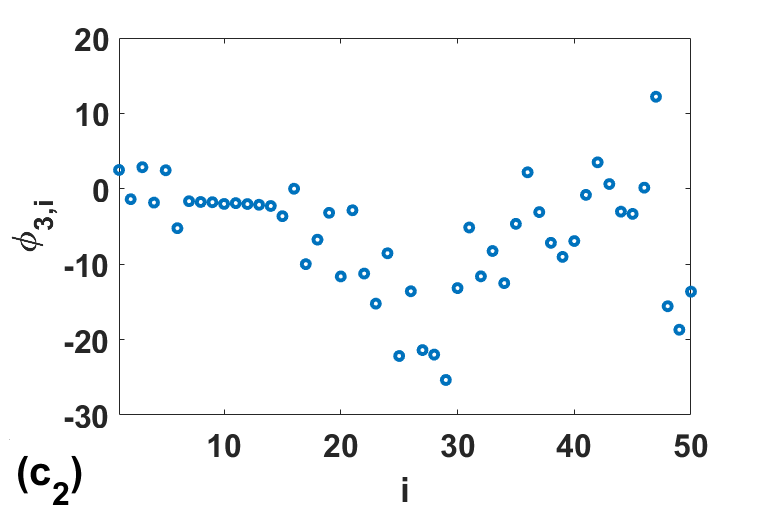}\\
    \includegraphics[width=0.5\textwidth]{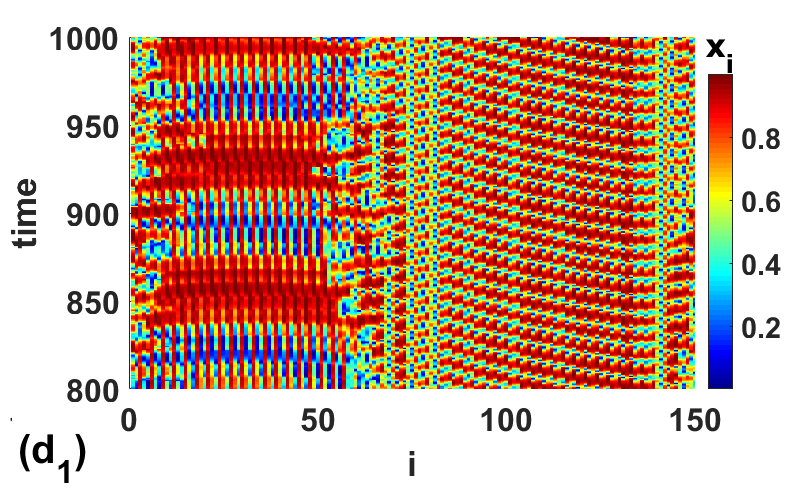}&
    \includegraphics[width=0.5\textwidth]{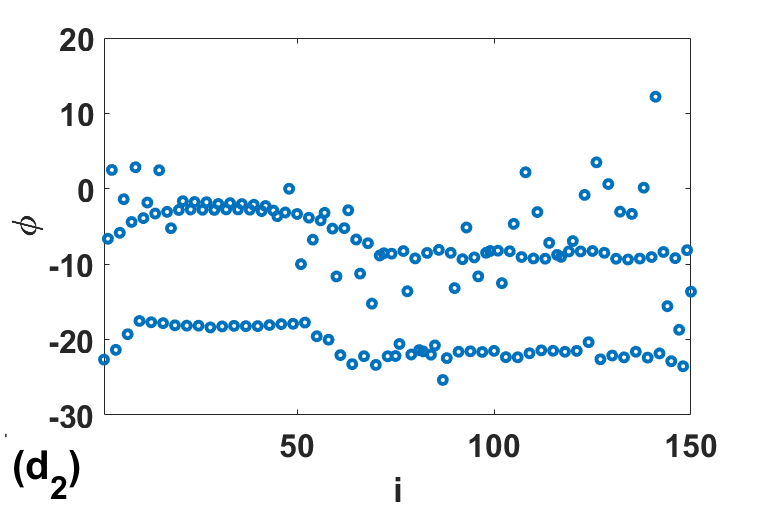}
    \end{tabular}
    \caption{The snapshots depict chimera states for the first ($(a_1)$, $(a_2)$), second ($(b_1)$, $(b_2)$), and third ($(c_1)$, $(c_2)$) neurons, respectively. Panels ($d_1$) and ($d_2$) offer a comprehensive, decomposed view of the internal dynamics across all N cells, corresponding to a total of 150 neurons. These illustrations were generated using the following parameters: $\sigma_D = 0.03$, $\sigma_P = 1.54$, $\sigma_I = 0.01$, and \(\epsilon = 1.5\).} 
    \label{snapshot2}
\end{figure}

To reinforce the overall results presented in Fig.\ref{oder1}, we provide in Fig.\ref{snapshot2} a more detailed analysis of selected time series and snapshots illustrating the collective behaviors formed by the first, second, and third neurons (see respectively Fig.\ref{snapshot2}($a_1, a_2$) for $k=1$, Fig.\ref{snapshot2}($b_1, b_2$) for $k=2$, and Fig.\ref{snapshot2}($c_1, c_2$) for $k=3$) with $\epsilon =1.5$, $\sigma_D=0.03$, $\sigma_P=1.54$ and $\sigma_I=0.01$. These figures highlight the similar behaviors (notably the chimera state) within the three neurons that make up each cell in the Hopfield network. Other behaviors like, Multichimera, coherence and synchronization could be observe in each neuron of the network as shown by the SI, DM and the order parameter in Fig.\ref{oder1}. 
In contrast, Fig.\ref{snapshot2}($d_1, d_2$) presents a decomposed view of the internal dynamics within each cell, enabling a visualization of neuronal dynamics within each of the N cells. Each cell, consisting of three neurons, results in a total of 3N neurons --i.e., 150 neurons as shown on the x-axis. This representation allows for the identification of a weak chimera state within the network when considering all 3N neurons. It is noteworthy that, while we observe an intercellular chimera state (i.e., the same variables within the N systems exhibit a chimera state), the entire 3N system does not exhibit a uniform chimera state. Instead, regions of coherence/synchronization and decoherence are not clearly visible as in first, second and third variables (see Fig. \ref{snapshot2}($a_1, c_2$)), highlighting that within the cells, coherence/synchronization is not always strongly maintained among the neurons of the same cell.

\begin{figure}[htp!]
\centering
  \begin{tabular}{cc}
    \includegraphics[width=0.33\textwidth]{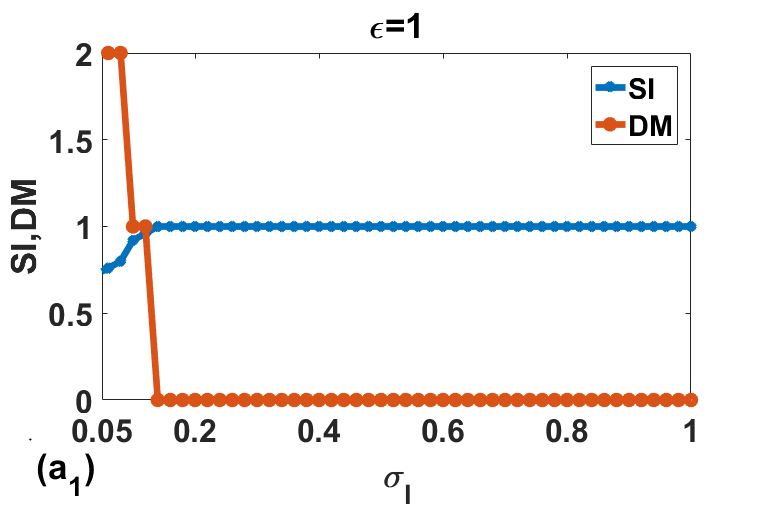}&
    \includegraphics[width=0.33\textwidth]{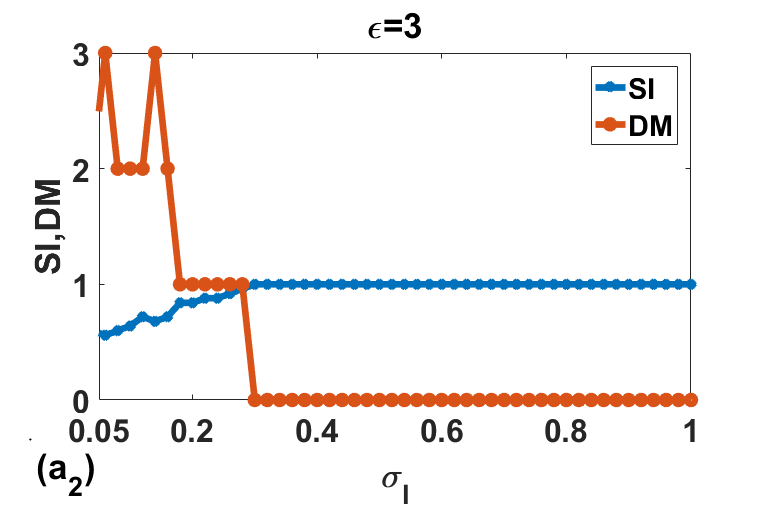}
    \includegraphics[width=0.33\textwidth]{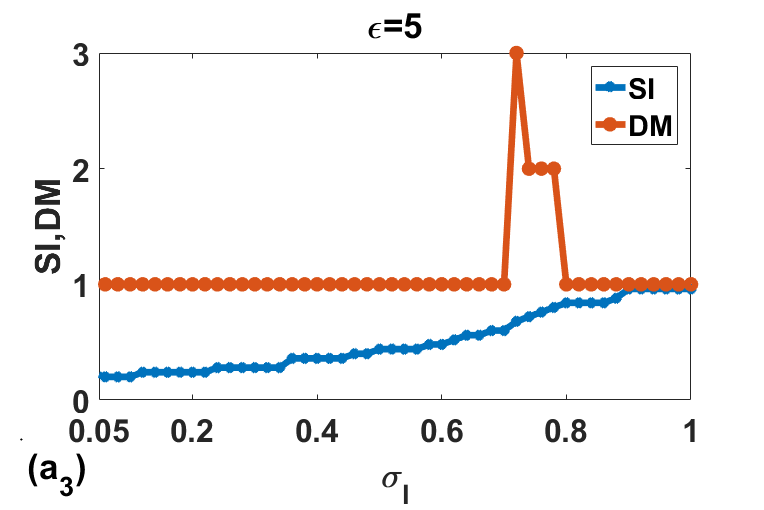}\\
    \includegraphics[width=0.33\textwidth]{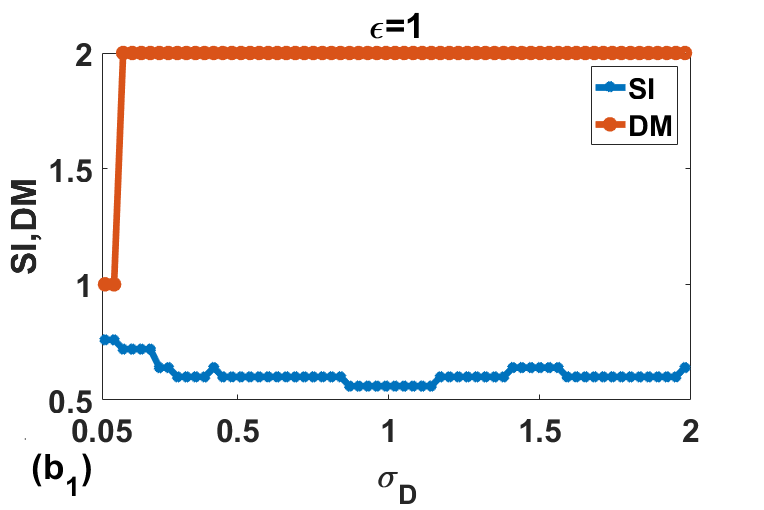}&
    \includegraphics[width=0.33\textwidth]{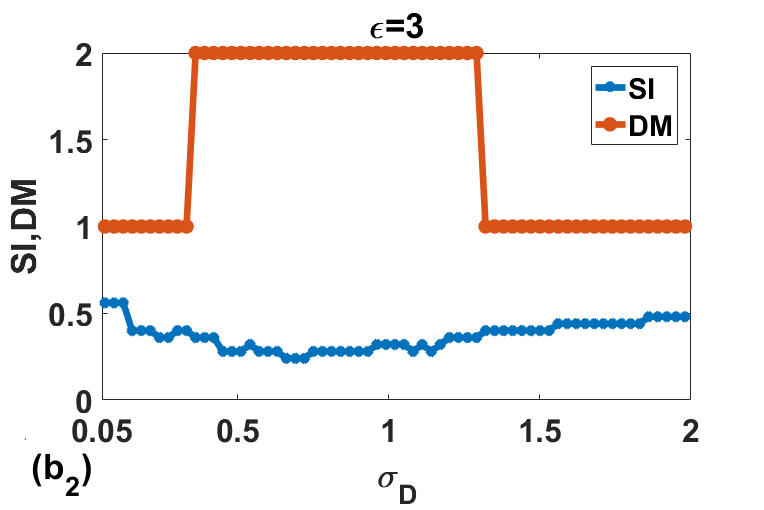}
    \includegraphics[width=0.33\textwidth]{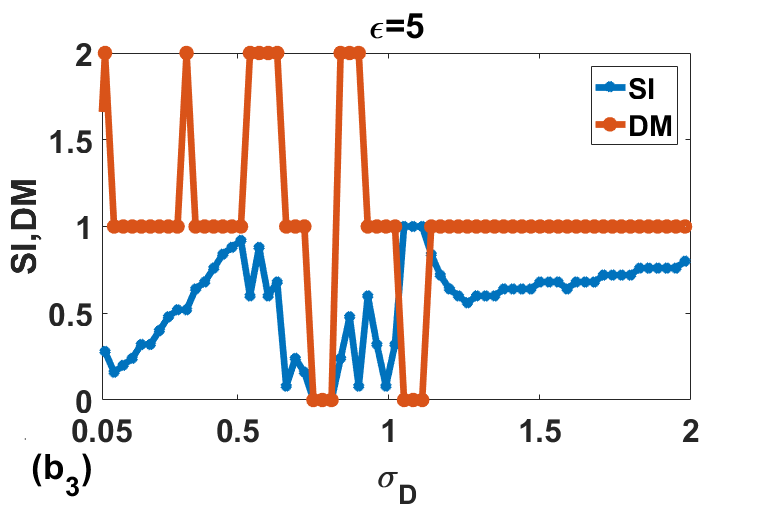}\\
    \includegraphics[width=0.33\textwidth]{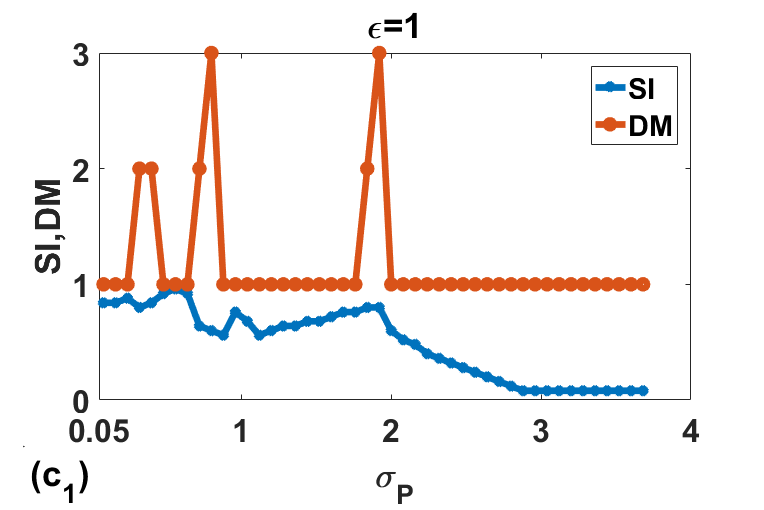}&
    \includegraphics[width=0.33\textwidth]{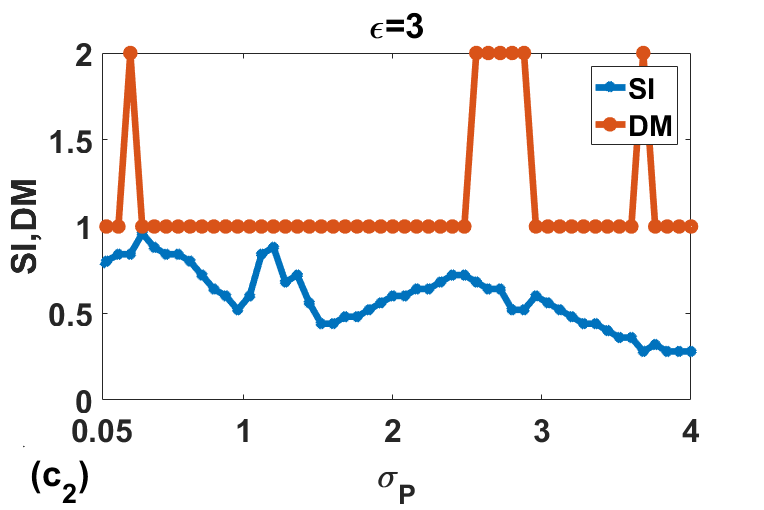}
    \includegraphics[width=0.33\textwidth]{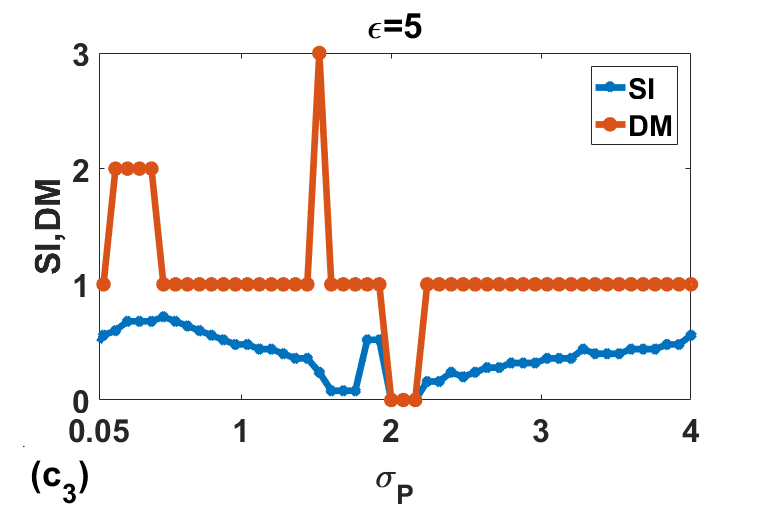}\\
\end{tabular}
    \caption{Impact of individual PID coupling coefficients on chimera state emergence, examined at three levels of intracellular coupling strength (\(\epsilon=1\), \(\epsilon=3\), and \(\epsilon=5\)). We report  the evolution of the Strength of Incoherence (SI) and Discontinuity Measure (DM) as a function of : \(\sigma_I\) (see panels $(a_1)$, $(a_2)$, and $(a_3)$); \(\sigma_D\) ( see panels $(b_1)$, $(b_2)$, and $(b_3)$) and \(\sigma_P\) (see panels $(c_1)$, $(c_2)$, and $(c_3)$). In each case, the other PID coupling coefficients are set to zero to isolate the effects of the individual coefficients.} 
    \label{oder1a}
\end{figure}

To clarify the role of each coupling coefficient in the PID coupling framework, we examine the individual effects of each coupling type in Fig.\ref{oder1a}. From top to bottom, the first row (see Fig.\ref{oder1a}($a_1-a_3$)) corresponds to the inductive coupling parameter, the second row (see Fig.\ref{oder1a}($b_1-b_3$)) to the derivative coupling parameter, and the third row (see Fig.\ref{oder1a}($c_1-c_3$)) to the proportional coupling parameter. In this analysis, each PID coupling parameter is varied independently over a specified range while the other parameters are held at zero. This examination is performed for three distinct values of the intra-cell coupling coefficient \(\epsilon=1\) for the first column, \(\epsilon=3\) second column, and \(\epsilon=5\) third column fo Fig.\ref{oder1a}. Each panel in Fig.\ref{oder1a} shows the evolution of the Strength of Incoherence (SI) and the Discontinuity Measure (DM).
The investigation of the effects of the inductive coupling coefficient \(\sigma_I\), depicted in the first row, shows that increasing intra-cell coupling strength expands the domain in which chimera states can emerge. Very low intra-cell coupling strength, however, tends to lead the system toward incoherent states. A similar conclusion applies to the derivative and proportional coupling parameters, with one exception: for low intra-cell coupling strengths, these parameters tend to drive the network toward a coherent state as coupling strength increases.
In summary, our findings demonstrate that the emergence of chimera states is influenced by the choice of coupling type. For example, depending on the value of the intra-cell coupling, inductive coupling may be unfavorable for achieving chimera states, instead promoting decoherence. Under the parameters and initial conditions considered in this study, as described above, derivative and proportional couplings appear to be the most suitable for supporting the formation of chimera states.

\begin{figure}[ht!]
\centering
  \begin{tabular}{cc}
\includegraphics[width=0.5\textwidth]{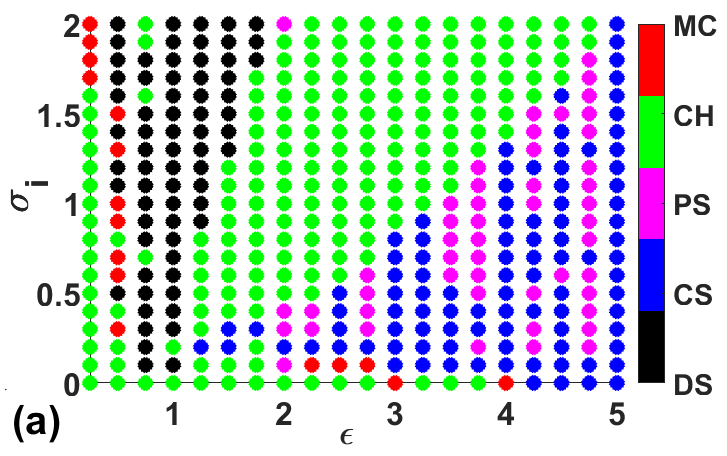}&
\includegraphics[width=0.5\textwidth]{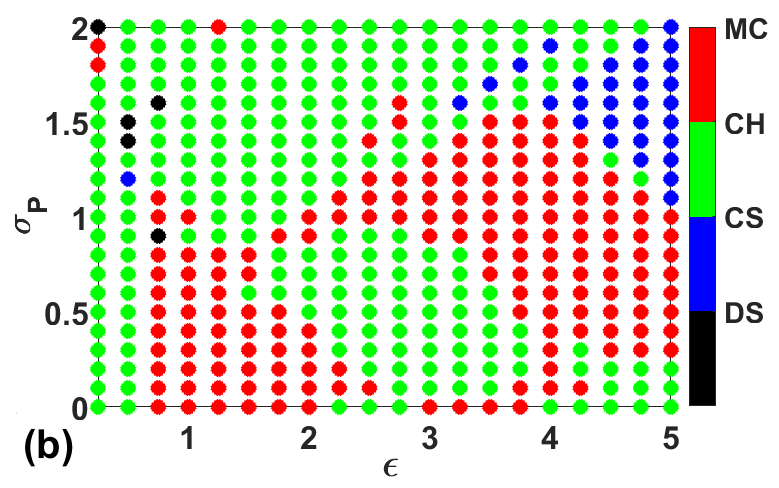}
\end{tabular}
\caption{Different collective behaviors exhibited by the network when simultaneously varying the parameters: (a) \((\epsilon, \sigma_I)\) with \(\sigma_P = 1.54\), and (b) \((\epsilon, \sigma_P)\) with \(\sigma_I = 0.01\). These results reveal five distinct states, each represented by a different color and acronym: black (DS) denotes desynchronization and traveling wave, blue (CS) indicates a coherent state, magenta (PS) signifies phase synchronization, green (CH) represents chimera states, and red (MC) corresponds to multichimera states. Other parameter \(\sigma_D = 0.3\).} 
\label{sidm}
\end{figure}

To gain a more comprehensive understanding of the various collective behaviors exhibited by this network, we analyze it from a broader perspective by simultaneously considering the evolution of two parameters: \(\epsilon\) for the intra-cell dynamics and \(\sigma_I\) and \(\sigma_P\) for the inter-cell dynamics, while keeping the other parameter constant (\(\sigma_D = 0.3\)). Using the tools mentioned earlier ($SI$, $DM$ and $r$), we represent in Fig.\ref{sidm} the different collective behaviors that the network can exhibit within the considered range of control parameters. Fig.\ref{sidm}(a) shows the behaviors for a simultaneous variation of the intra-cell coupling strength $\epsilon$ and the inter-cell coupling strength $\sigma_I$, with $\sigma_P$ and $\sigma_D$ held constant. Fig.\ref{sidm}(b) depicts the behaviors for a simultaneous variation of the intra-cell coupling strength $\epsilon$ and the inter-cell coupling strength $\sigma_P$, with $\sigma_D$ and $\sigma_I$ held constant.

\begin{figure}[htp]
\centering
  \begin{tabular}{cc}
    \includegraphics[width=0.5\textwidth]{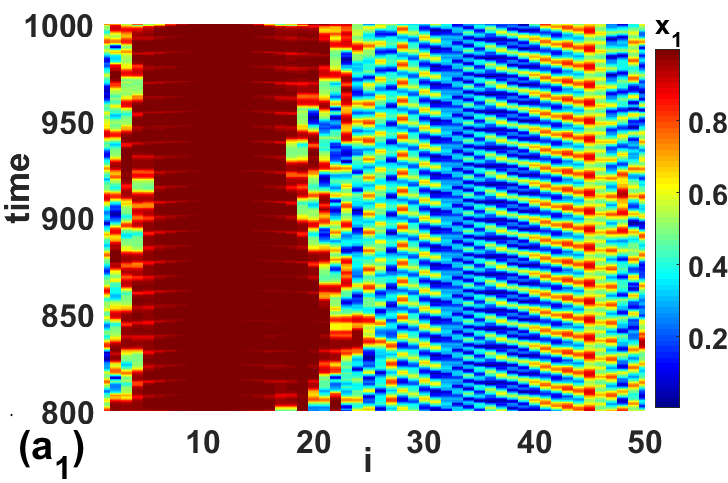}&
    \includegraphics[width=0.5\textwidth]{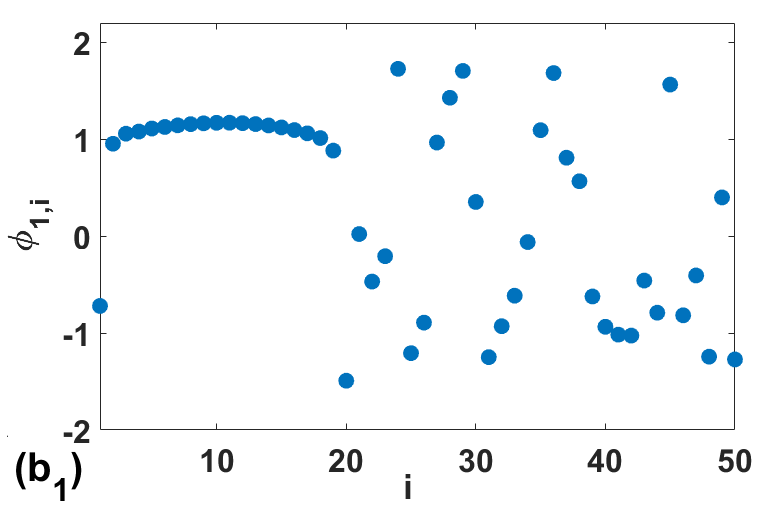}\\
    \includegraphics[width=0.5\textwidth]{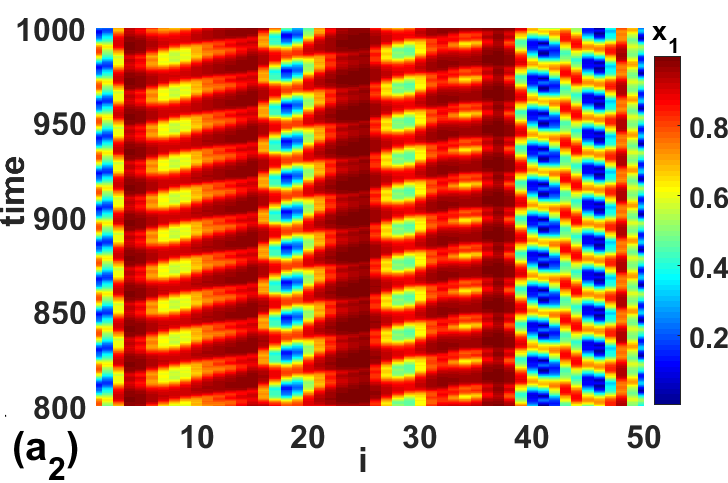}&
    \includegraphics[width=0.5\textwidth]{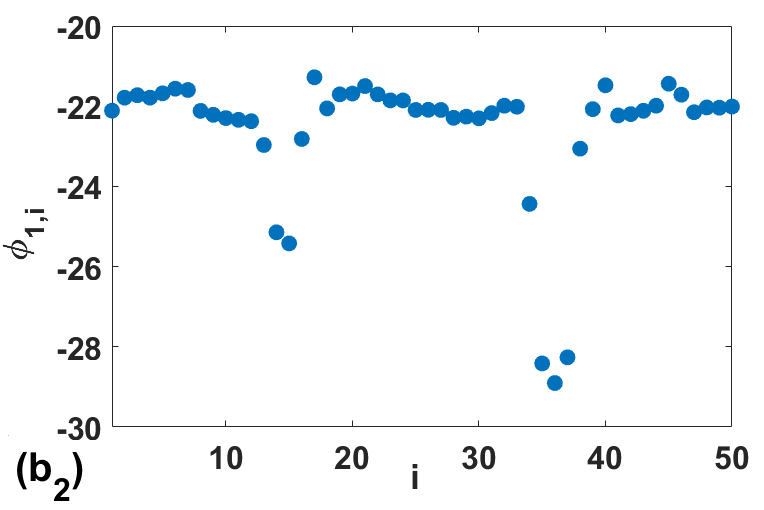}\\
    \includegraphics[width=0.5\textwidth]{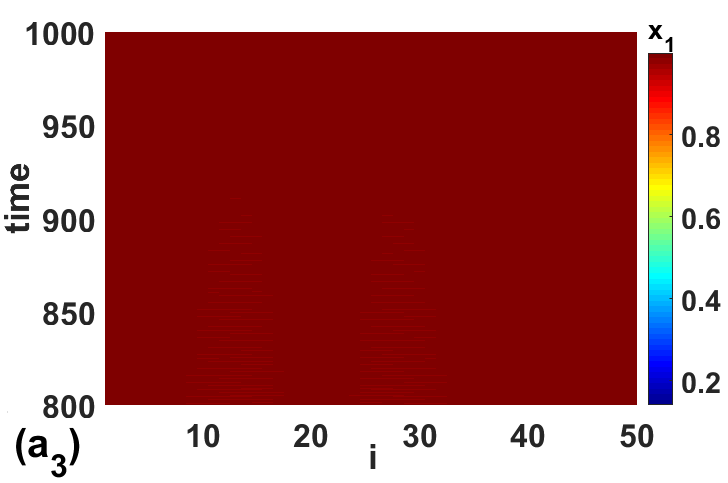}&
    \includegraphics[width=0.5\textwidth]{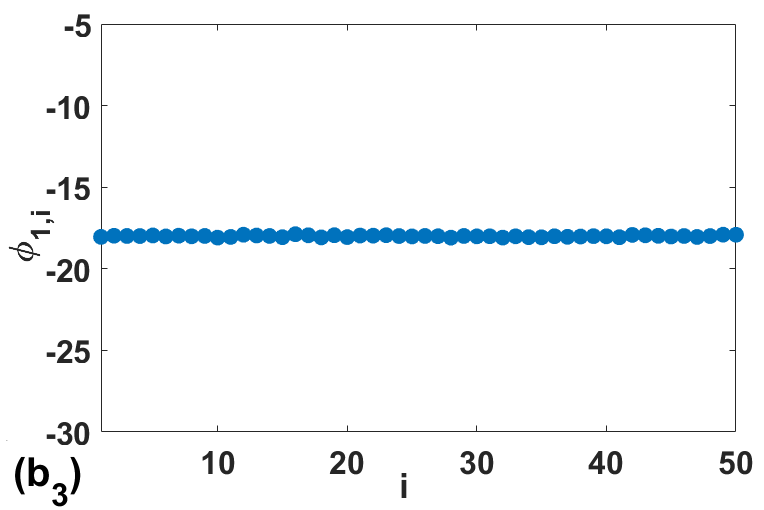}\\
    \includegraphics[width=0.5\textwidth]{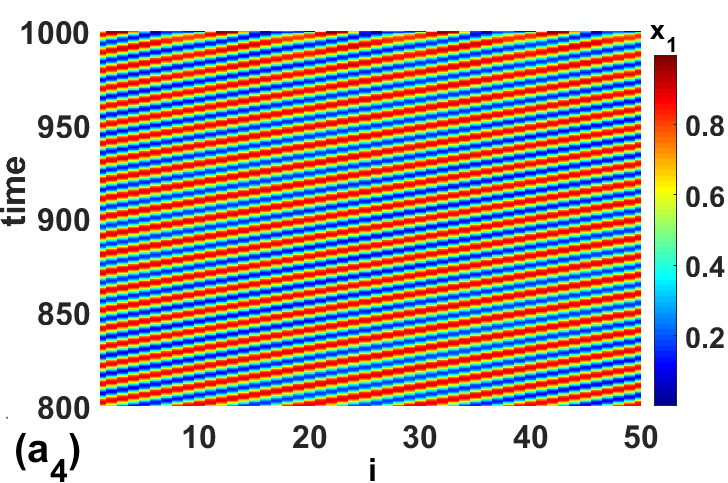}&
    \includegraphics[width=0.5\textwidth]{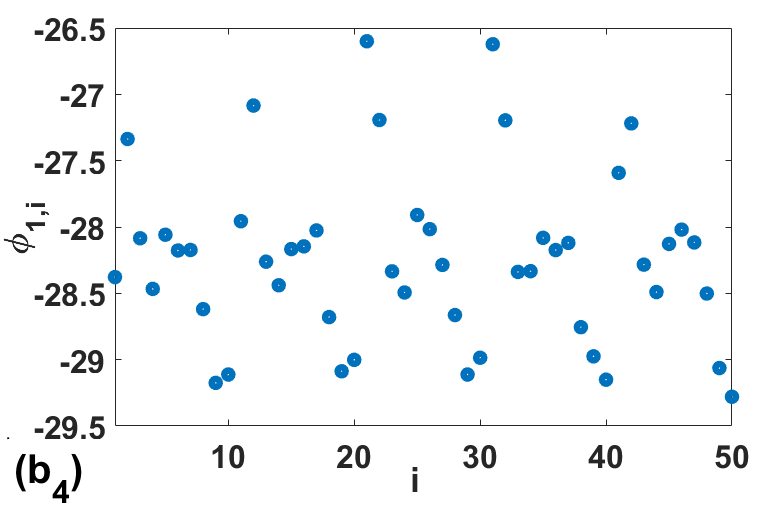}
    \end{tabular}
    \caption{Snapshots highlight different chimera states depicted in Fig.\ref{sidm}. These illustrations are obtained with \(\sigma_D = 0.3\) and \(\sigma_I = 0.01\), except for the last row (\((a_4)\) and \((b_4)\)) where \(\sigma_I\) is 1.13. Specifically, \((a_1)\) and \((b_1)\) correspond to a chimera state in region \(CH\) (\(\epsilon = 2\), \(\sigma_P = 1.54\)); \((a_2)\) and \((b_2)\) represent a multichimera state in region \(DM\) (\(\epsilon = 4\), \(\sigma_P = 1.0\)); \((a_3)\) and \((b_3)\) depict a coherent state in region \(CS\) (\(\epsilon = 3.9\), \(\sigma_P = 1.54\)); and \((a_4)\) and \((b_4)\) illustrate a traveling wave state in region \(DS\) (\(\epsilon = 1.3\), \(\sigma_P = 1.54\)).} 
    \label{snapshot}
\end{figure}

In these two scenarios, the results highlight five distinct domains, each identifiable by a specific color and acronym: black (DS) for desynchronization and traveling wave, blue (CS) for coherent state, magenta (PS) for phase synchronization, green (CH) for chimera states, and red (MC) for multichimera states. Fig.\ref{sidm}(a) is obtained for \(\sigma_P = 1.54\), a value that estimates the significant resistive influence. In the absence of the inductive part of the coupling (\(\sigma_I = 0\)), chimera states are observed for low values of the intra-cells coupling strength. However, as the intra-cells coupling strength increases, coherent states form. Conversely, when \(\sigma_I\) increases, an increase in the intra-cells coupling strength promotes the formation of chimera states. In general, a simultaneous increase in \(\epsilon\) and \(\sigma_I\) predominantly leads to four behaviors: desynchronization, coherence, phase synchronization, and chimera states. Fig.\ref{sidm}(b) illustrates that with \(\sigma_I = 0.01\) and \(\sigma_D = 0.3\), the simultaneous variation of \(\epsilon\) and \(\sigma_P\) predominantly favors the formation of multichimera states, chimera states, and coherent states. This suggests that even a slight inductive coupling (\(\sigma_I\)) in combination with resistive coupling (\(\sigma_P\)) can lead to complex and diverse behaviors. The interplay between \(\epsilon\) and \(\sigma_P\) appears to create conditions conducive to the emergence of these intricate states, highlighting the sensitivity of the network dynamics to these parameters. To illustrate these different behaviors, Fig.\ref{snapshot} serves as a visual representation, showcasing both the temporal evolution of the local order parameter and snapshots depicting the average phase of each cell over time. These visualizations are crucial for assessing coherence or phase synchronization at the cellular level. The local order parameter, defined by Eq.\ref{local} and referenced in previous studies \cite{wan2017self, wolfrum2011spectral}, plays a pivotal role in quantifying the degree of synchronization among cells. It offers a precise measure to distinguish between coherent states where cells exhibit synchronized behavior and phase synchronization where cells exhibit synchronized phase but not necessarily amplitude.
\begin{equation}\label{local}
  {L_i} = \left| {\frac{1}{{2q}}\sum\limits_{\left| {i - j} \right| \le q} {{e^{\ell{\phi _j}}}} } \right|
\end{equation}
In this equation, \( q \) represents the number of neighbors to the left and right of a node \( i \), \( \ell^2 = -1 \), and the geometric phase \( \phi_k \) is defined by the expression:
\begin{equation}\label{phase}
  {\phi _j} = \arctan \left( {\frac{{img({x_j})}}{{real({x_j})}}} \right)
\end{equation} 
A value of \(L_i\) close to 1 indicates that the phases of neurons in this node are aligned, suggesting strong phase synchronization. If \(L_i\) is significantly less than $1$, it indicates that the phases of neurons in node \(j\) are distributed randomly or desynchronized.

The first column of Fig.\ref{snapshot} (see \(a_1\) - \(a_4\)) presents the local order parameter, while the second column (see \(b_1\) - \(b_4\)) shows the corresponding snapshots of the time-averaged phase \(\phi_j\) of each node to clearly highlight the hidden dynamics. The values of \(L_j\) (see Eq.\ref{local}) for the first neurons (\(k=1\)) of the \(j^{th}\) node in the \((j, x^1_j)\) space are indicated by the colormap in the left panel (Figs.\ref{snapshot}.(($a_1$) to ($a_4$))). Fig.\ref{snapshot} provides a comprehensive visual representation of four different dynamical behaviors observed within the network. Chimera State (panels \(a_1\) and \(b_1\)): This state shows a coexistence of synchronized and desynchronized domains within the network, where certain groups of nodes synchronize their activity while others remain unsynchronized. This phenomenon is visually highlighted by distinct patterns in both the local order parameter (\(L_j\)) and the snapshots of average phase \(\phi_j\) across nodes. The multichimera State (panels \(a_2\) and \(b_2\)): Here, the network exhibits multiple coexisting domains of synchronized and desynchronized behavior, often forming complex spatial patterns. The local order parameter and phase snapshots depict these intricate patterns, indicating a higher level of complexity compared to standard chimera states. The synchronous State (panels \(a_3\) and \(b_3\)). This state demonstrates complete Phase synchronization across all nodes in the network, where all nodes exhibit identical phases. The local order parameter approaches unity, indicating strong synchronization, while phase snapshots show uniform phase distribution across nodes. The traveling wave State (panels \(a_4\) and \(b_4\)): In this state, localized regions of synchronized and desynchronized dynamics move through the network over time. This dynamic behavior is captured by observing changes in the local order parameter and the spatial evolution of phase snapshots. These visual representations not only illustrate the diverse behaviors inherent in complex dynamical systems but also provide valuable insights into how system parameters (\(\epsilon\), \(\sigma_P\), \(\sigma_I\) and  \(\sigma_D\).) influence the emergence and stability of these states. Such detailed analysis helps to show the impact of the PID coupling on the network dynamics.

\subsection{Dependence on the initial conditions}\label{sec4c}
\subsubsection{Multistability}\label{sec4s}

\begin{figure}[htp]
\centering
  \begin{tabular}{cc}
    \includegraphics[width=0.5\textwidth]{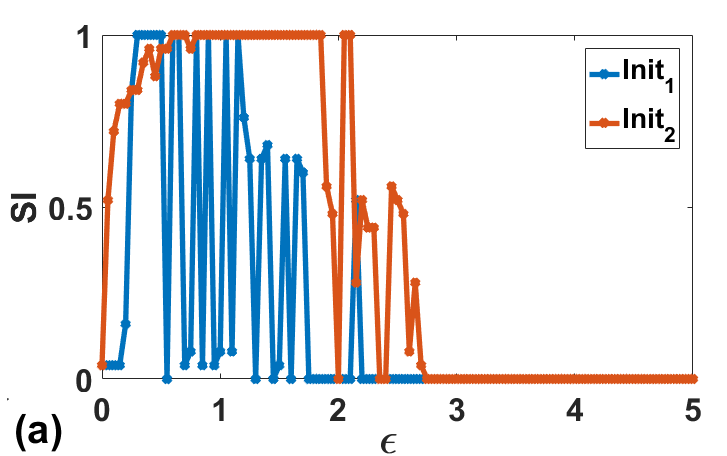}
    \includegraphics[width=0.5\textwidth]{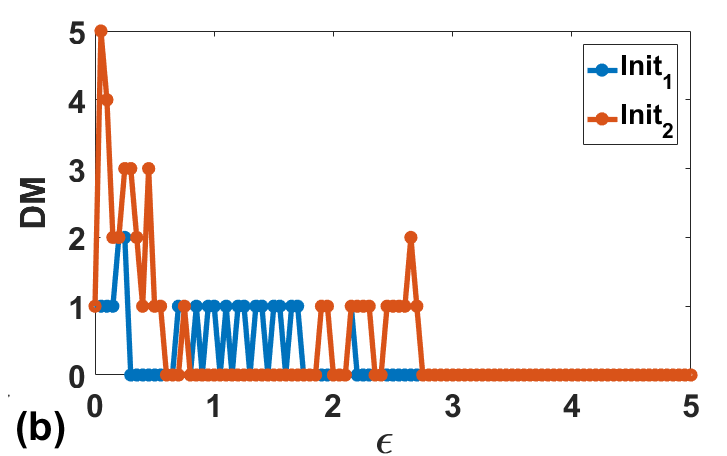}
    \end{tabular}
    \caption{We investigate the system dynamics under two distinct initial conditions for \(x_{4,i}\), denoted as \(Init_1\) and \(Init_2\). Panel (a) presents the temporal evolution of the strength of incoherence, while panel (b) illustrates the discontinuity measure. These analyses are conducted under fixed parameters \({\sigma_D}=0.3\), \({\sigma_P}=1.4\), and \({\sigma_I}=0.7\).}\label{snapshot1}
\end{figure}

In this subsection, we explore the influence of initial conditions on the observed behaviors named multistability. Multistability stands as a pivotal property within nonlinear interactive systems, offering profound insights into the dynamics of complex phenomena. It characterizes systems where even minor disturbances can trigger transitions to entirely new states that exhibit distinct characteristics from their initial conditions \cite{feudel1997multistability,sharma2014controlling,sharma2015control}. This property is indispensable for interpreting a myriad of natural and pathological processes across different scales of biological organization, encompassing phenomena like the cell cycle and immune memory. We define two groups of initial conditions labeled \(Init_1\) and \(Init_2\) randomly generated with uniform distributions in the interval $\left[0,1\right]$ for the state variable \(x_{4,i}\) consequence of the PID coupling. Fig.\ref{snapshot1} illustrates the dynamics of the Strength of Incoherence (\(SI\)) in panel (a) and the Discontinuity Measure (\(DM\)) in panel (b), evaluated using the first neurons within each cell.

Across these different initial conditions, our investigation reveals consistent dynamic states observed in the system, albeit appearing at quantitatively different values for the same control parameter settings. For instance, within the range \(1 \leq \epsilon \leq 1.8\), clear coexistence is observed: the chimera state (\(SI \leq 1\) and \(DM = 1\), indicated by the blue curve) and total incoherence (\(SI = 1\) and \(DM = 0\), represented by the red curve). These results show the multistability of the collective behaviors for certain values of the intra-cells coupling strength.

\subsubsection{ Control of the multistable behavior. }\label{sec4}

Let's delve into strategies for controlling multistability, focusing particularly on a linear augmentation scheme \cite{njitacke2021window,sharma2011targeting}. This approach aims to steer multistable systems towards monostable states through systematic adjustments. By augmenting the system with linear components, researchers seek to stabilize dynamics and enhance predictability in system behavior. This pursuit is pivotal in practical applications where stability and reproducibility are paramount, such as in designing robust biological circuits or optimizing chemical processes. Moreover, understanding and manipulating multistability not only advances theoretical models but also drives experimental innovations, facilitating breakthroughs in fields spanning biophysics to engineering.

Let us Consider the previous Eq.\ref{eq1}, the augmented system is given by:

\begin{equation}\label{multi1}
\left\{ \begin{array}{l}
{\dot x_{k,i}} =  - {x_{k,i}} + \epsilon \sum\limits_{l = 1}^n {{w_{kl}}\tanh \left( {{\beta _l}{x_{l,i}}} \right)} + {I_{ij}} + \alpha H\left( y_i \right) \\
  \dot y_i = -s{y_i}-\alpha \left( {x_{3,i}}-b \right)\\
\end{array} \right.\,
\end{equation}

The variable \( y \), which is one-dimensional, characterizes the dynamics of the linear system. In this context, \( s \) serves as the decay parameter, and \( b \) is the control parameter of the augmented system \cite{sharma2011targeting}. Here, \( \alpha \) represents the strength of the feedback between the neuron and the linear system, with \( H(y_i) = H(0, 0, y_i) \). Utilizing Eq.\ref{multi1}, we derive the network of the augmented system, which can be expressed as:
\begin{equation}\label{multi2}
  \left\{ \begin{array}{l}
\dot x_{1,i} =  - x_{1,i} + \epsilon (2\tanh \left( {{\beta _1}x_{1,i}} \right) - 1.2\tanh \left( {{\beta _2}x_{2,i}} \right) + 0.48\tanh \left( {{\beta _3}x_{3,i}} \right))\\
\dot x_{2,i} =  - x_{2,i} + \epsilon (3.6\tanh \left( {{\beta _1}x_{1,i}} \right) + 1.7\tanh \left( {{\beta _2}x_{2,i}} \right) + 1.076\tanh \left( {{\beta _3}x_{3,i}} \right))\\
\dot x_{3,i} = x_{4,i}\\
\dot x_{4,i} = \sum\limits_j {M_{ij}^{ - 1}{F_j^1}}\\
\dot y_i = -s{y_i} - \alpha \left( {x_{3,i}-b} \right))
\end{array} \right.\,
\end{equation}

With $M$ the matrix defined by Eq.\ref{eq7} and the term ${F_j^1}$ by Eq.\ref{eq6a}:
\begin{equation}\label{eq6a}
  \begin{array}{l}
{F_j^1} =  - x_{4,j} + \alpha \dot y_j- \frac{{9\epsilon {\beta _1}\dot x_{1,j}}}{{{{\left( {\cosh \left( {{\beta _1},x_{1,j}} \right)} \right)}^2}}} + \frac{1}{{2q}}\sum\limits_{\ell = j - q,\,j \ne \ell}^{j + q} {\left( {{\sigma_P}\left( {x_{4,\ell} - x_{4,j}} \right) + {\sigma_I}\left( {x_{4,\ell} - x_{3,j}} \right)} \right)}
\end{array}
\end{equation}

\begin{figure}[htp]
\centering
  \begin{tabular}{cc}
    \includegraphics[width=0.5\textwidth]{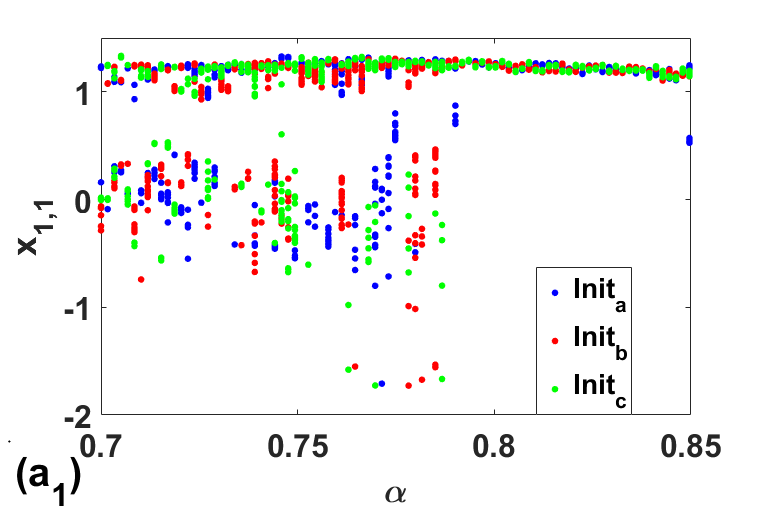}&
    \includegraphics[width=0.5\textwidth]{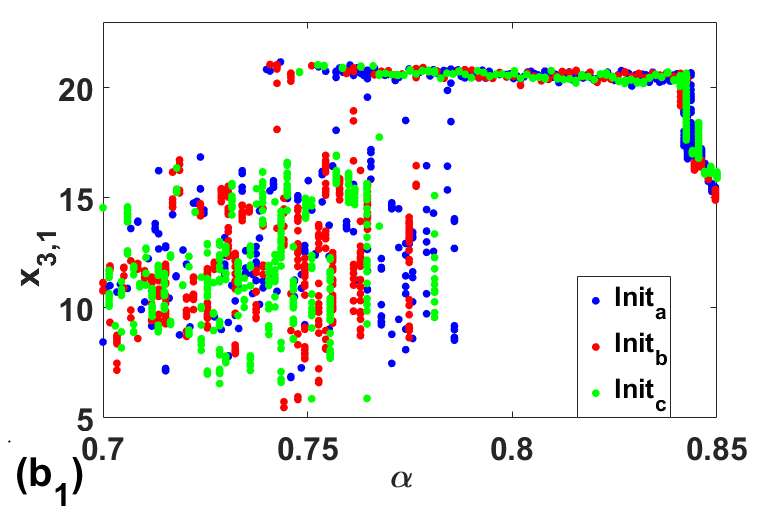}\\
    \includegraphics[width=0.5\textwidth]{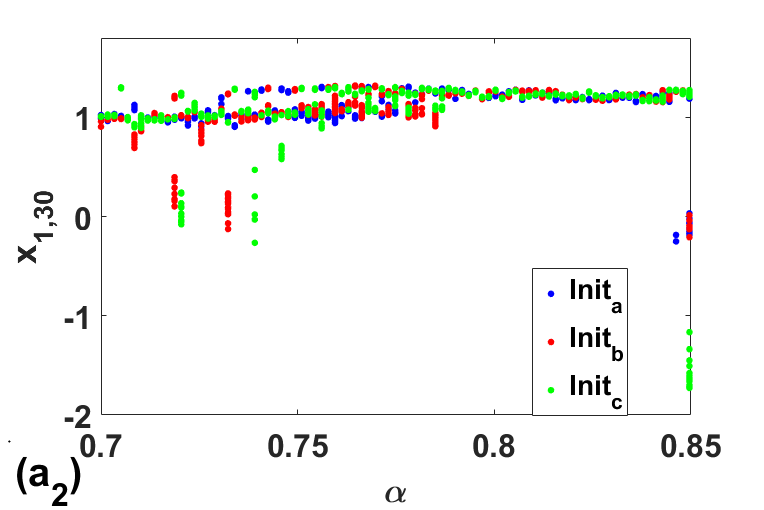}&
    \includegraphics[width=0.5\textwidth]{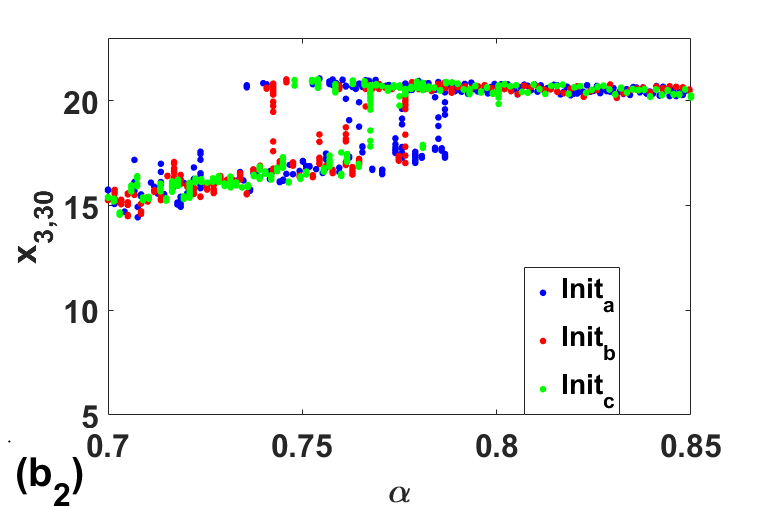}\\
    \includegraphics[width=0.5\textwidth]{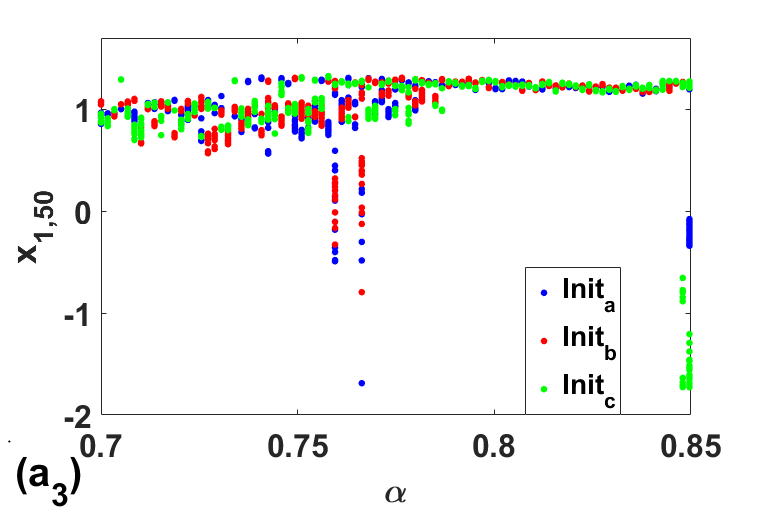}&
    \includegraphics[width=0.5\textwidth]{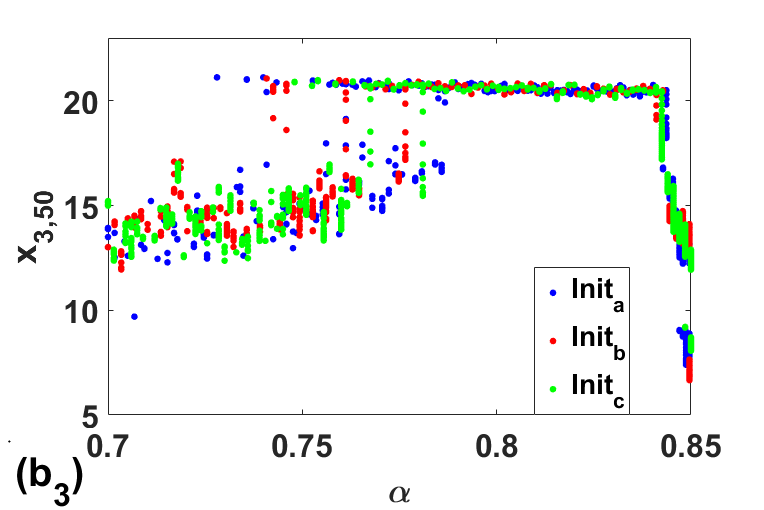}
    \end{tabular}
    \caption{We report the bifurcation diagrams for the first and third neurons in the 1st, 30th, and 50th cells of the network, each under three distinct initial conditions: \( Init_a \) (blue), \( Init_b \) (red), and \( Init_c \) (green), plotted as a function of feedback strength (\( \alpha \)). The parameters used are \( \sigma_I = 0.01 \), \( \sigma_P = 1.54 \), \( \sigma_D = 0.2 \), \( b = 8.9759 \), \( s = 0.5 \), and \( \epsilon = 1 \).}
    \label{bifurc1}
\end{figure}

To assess the stability of the network’s results, we present in Fig.\ref{bifurc1} the bifurcation diagrams for the first and third neurons in the first cell (see Fig.\ref{bifurc1}($a_1, b_1$)), the 30th cell (see Fig.\ref{bifurc1}($a_2, b_2$)), and the 50th cell (see Fig.\ref{bifurc1}($a_3, b_3$)). For this analysis, we consider three different initial conditions: \( Init_a \) (blue), \( Init_b \) (red), and \( Init_c \) (green). These initial conditions apply to the state variable \( x_{4,i} \), introduced into the system via PID coupling, and vary within the interval \([0, 1]\). The bifurcation diagram illustrates the system’s response to changes in the feedback strength (\( \alpha \)) between the neuron and the linear system, represented by the variable \( y_i \). The figure illustrates in all cases the superposition of three bifurcation diagrams, each corresponding to a different set of initial conditions. This superposition highlights the phenomenon of coexisting bifurcations, indicating that multiple stable states can exist for the same synaptic weights but different initial conditions \cite{njitacke2021window}. This behavior is significant as it underscores the system's sensitivity to initial conditions and the potential for diverse dynamic outcomes. We note that a similar behavior is observed when using the second neuron within the cell. The parameters used in this analysis are: \( \sigma_I = 0.01 \); \( \sigma_P = 1.54 \); \( \sigma_D = 0.2 \); \( b = 8.9759 \); \( s = 0.5 \); and \( \epsilon = 1 \).

\begin{figure}[htp]
\centering
  \begin{tabular}{c}
    \includegraphics[width=1\textwidth]{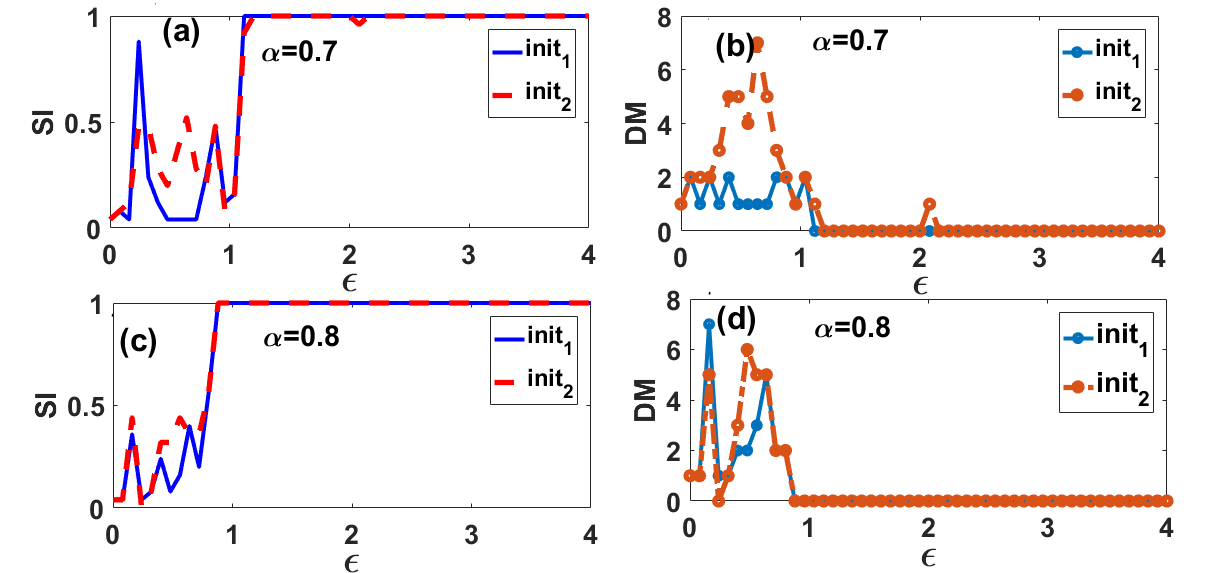}
  \end{tabular}
\caption{The effect of control on the collective behaviors exhibited by the network, as functions of the intracellular coupling force \( \epsilon \), is analyzed for two values of the coupling force within the linearized system (\( \alpha \)). These dynamics are illustrated for two distinct sets of initial conditions. The parameter values are set as follows: \( \sigma_I = 0.3 \); \( \sigma_P = 1.4 \); \( \sigma_D = 0.7 \); \( b = 8.9759 \); and \( s = 0.5 \). }\label{bifurc2}
\end{figure}

In the diagrams of Fig.\ref{bifurc2}, we study the influence of the coefficient \( \alpha \) on the emergence of collective behavior as a function of two initial conditions. We varied the intracellular coupling coefficient for two different values of the coefficient \( \alpha \), corresponding to the zones of multistability and stability as shown in the previous Fig.\ref{bifurc1}. It turns out that the higher the coefficient \( \alpha \), the more the system tends towards a state of disorder. We can therefore deduce that phenomena such as synchronization and chimera states disappear when the \( \alpha \) instability coefficient increases. This behavior is due to the fact that the greater the value of the coupling coefficient with the linearized \( \alpha \) system, the greater the disturbance to the system. 

Conversely, these figures for the two values \( \alpha = 0.7 \) and \( \alpha = 0.8 \) show practically the same behavior for different initial conditions. This demonstrates that the applied control effectively transitions the system from a multistable behavior to a monostable behavior. By stabilizing the system across varying initial conditions, the control mechanism proves robust in eliminating multistability, thus ensuring consistent collective behavior regardless of initial state variability. This highlights the effectiveness of the linear augmentation scheme in managing complex dynamics within the network.

\section{Conclusion}\label{sec5}
This paper investigates the collective behaviors in a cell network composed of three Hopfield neurons in each cell of the network and coupled using PID (Proportional-Integral-Derivative) coupling. PID coupling could be used in neural network models to mimic the electrical and chemical properties of synapses, enabling the regulation of complex dynamic behaviors like synchronization, additionally, it is applied in technological controls to manage electrical and magnetic properties, echoing how synapses modulate bioelectric signals \cite{zhang2020multi}.
The dynamics of this model reveal the emergence of synchronization between cells, chimera states, multichimera states, and traveling chimera states. The study identifies and characterizes these behaviors using several metrics: the order parameter measures the degree of synchronization, the local order parameter distinguishes between chimera states and synchronized states, the strength of incoherence quantifies the extent of incoherent and coherent behavior (chimera states), and the discontinuity measure detects the presence of multichimera states.

Furthermore, the model exhibits multistability, where the system can reside in multiple stable states under the same parameters and different initial conditions. By applying a linear augmentation scheme, the study proposes a control mechanism that transitions the system from a multistable state to a monostable state. This control method is crucial for practical applications, ensuring the stability and predictability of the system. The nature of the PID coupling and the resulting dynamic behaviors demonstrate that this model possesses a rich and complex dynamical repertoire, making it particularly valuable for studying brain diseases, where understanding synchronization and desynchronization processes can provide insights into conditions such as epilepsy and Parkinson's disease. The findings underscore the potential of using this model as a tool for exploring neurological disorders and developing effective interventions.

\section*{Acknowledgements}
TN thanks the ``Reconstruction, Resilience and Recovery of Socio-Economic Networks'' RECON-NET EP\_FAIR\_005 - PE0000013 ``FAIR'' - PNRR M4C2 Investment 1.3, financed by the European Union – NextGenerationEU for the partial support. PL, TN, SJK, NMV and MoCLiS research group thanks ICTP for the equipment donation under the letter of donation Trieste $12^{th}$ August 2021. PL thanks thanks the support of the German Academic Exchange Service (DAAD) for funding his visit at the Potsdam Institute for Climate Impact Research (PIK) under the Grant Number (91897150).
	
\appendix
\section{Order parameter}\label{appA}

The order parameter, introduced by Kuramoto and Battogtokh \cite{kuramoto2002coexistence}, is a valuable tool for analyzing phase synchronization in coupled systems and networks. Its calculation requires knowledge of the phase of each individual system. To calculate this phase, we
consider an arbitrary time signal $ s\left( \tau  \right)$  and $ \tilde s\left(
\tau  \right) $ being its Hilbert transform. We have:
\begin{equation}\label{o1}
\psi \left( \tau  \right) = s\left( \tau  \right) + \ell\tilde s\left( \tau
\right) = R\left( \tau  \right){e^{\ell\varphi \left( \tau  \right)}}
\end{equation}
where $ R\left( \tau  \right)$ is the amplitude and $ \varphi \left( \tau
\right) $ the phase of the variable $ s\left( \tau  \right)$. If we denote by $
{\varphi_i} \left( \tau  \right) $ the instantaneous phase, then it can be
determined by:
\begin{equation}\label{o2}
{\varphi _i}\left( \tau  \right) = {\tan ^{ - 1}}\left( {\frac{{{{\tilde
s}_i}\left( \tau  \right)}}{{{s_i}\left( \tau  \right)}}} \right)
\end{equation}
The order parameter for a system with $ N $ oscillators is expressed as:
\begin{equation}\label{o3}
  r = \frac{1}{N}\sum\limits_{i = 1}^N {{e^{\ell{\varphi_i}}}}
\end{equation}
Where $ {\ell^2} =  - 1 $. Phase synchronization is effective when the value for $ r \to 1 $ and when $ r \to 0 $, the network is completely desynchronized.

\section{Strength of Incoherence (SI) and Discontinuity Measure (DM)}\label{appB}

The Strength of Incoherence (SI) is a statistical metric introduced by Gopal et al. \cite{gopal2014observation} to characterize incoherent, chimera, and coherent states within networks. This metric is defined by Eq.\ref{is}.

\begin{equation}\label{is}
  SI = 1 - \frac{1}{M}\sum\limits_{m = 1}^M {{s_m}}
\end{equation}
Where $m=1, 2, ... , M$ with $M$ the number of groups of equal length $n = N/M$ and ${s_m} = \Theta \left( {\delta  - \sigma \left( m \right)} \right)$. $\Theta \left( . \right)$ is the Heaviside step function, and $\delta$ is a small predefined threshold. $\sigma \left( m \right)$ is the local standard deviation and its expression is given by the following equation.

\begin{equation}\label{sigma}
  \sigma \left( m \right) = {\left\langle {\sqrt {\frac{1}{n}\sum\limits_{j = n\left( {m - 1} \right) + 1}^{nm} {{{\left[ {{z_j} - \bar z} \right]}^2}} } } \right\rangle _t}
\end{equation}
with \( \bar{z} = \frac{1}{N} \sum_{i=1}^N z_i \). The \( z_i \) are the dynamic variables of differences, defined as \( z_i = z_{i+1} - z_i \), and \( \langle \cdot \rangle_t \) denotes a time average. A Strength of Incoherence (SI) value of 0 indicates a coherent state, a value of 1 corresponds to an incoherent state, and values between 0 and 1 represent chimera states (and multichimera states).

The Discontinuity Measure (DM) is based on the distribution of \( s_m \) and is used to differentiate between chimera and multi-chimera states. The DM value is 0 in cases of either total coherence or total decoherence. A value of 1 indicates a simple chimera state with a single coherence domain. Integer values greater than 1 but less than half the number of nodes in the network correspond to multichimera states \citep{gopal2014observation}.

\begin{equation}\label{eqdm}
  DM  = \frac{1}{2}\sum\limits_{i = 1}^M {\left| {{s_i} - {s_{i + 1}}} \right|} %
\end{equation}

\bibliographystyle{unsrt} 
\bibliography{ref}

\end{document}